*Pulkovo Observatory of Russian Academy of Science, Saint-Petersburg*

## Introduction

The 2009 WZ104 asteroid was discovered with 68-cm Schmidt Catalina Sky Survey telescope of Mount Lemmon Observatory in Arizona (USA) in 25 of November 2009. The observations of the new object were confirmed with 60-cm reflector of Schiaparelli observatory (Italy) and 35-cm Schmidt – Cassegrain telescope of Alter Saltzberg observatory (Vienna, Austria). The asteroid was attributed as potentially hazardous near Earth asteroid. The minimal orbit intersection distance is 0.0304 a.e.

The observations of the 2009 WZ104 asteroid were continued with MTM-500M 50−cm meniscus telescope and ZA-320M 32-cm mirror astrograph of Pulkovo observatory (Russia) since late 2009. The observations were made in the framework of NEOs investigation program. The following tasks were established for the investigation: to get astrometric and photometric (in *BVRI* bands) series of observations, to improve the asteroid orbit on the base of the observational data, to determine its taxonomic class and absolute magnitude, to investigate the dynamics of its rotation, to estimate its physical parameters. The asteroid was attributed to Aten group among NEAs.

## Observations and analysis

The 2009 WZ104 asteroid was observed with telescopes throughout the world since November of 2009 to February of 2010 on the arc of 92°. The 1076 observational points were made in 26 observatories. The Fig. 1 shows the diagram of observation amount for different observatories. The Pulkovo observatory portion is about 74%.

Our observations were made on 8–12 of December 2009 with MTM-500M meniscus telescope (Kulish et al, 2009) at Mount astronomical station (Northern Caucasus) and on 16–21 of December 2009 with ZA-320M mirror astrograph (Devyatkin et al., 2009) at Pulkovo observatory (Saint-Petersburg). Astrometric and photometric processing of the observations was made with use of APEX-II software (Devyatkin et al., 2007; Devyatkin et al., 2010). The obtained coordinates and magnitudes (in the integral band of ZA-320M) of 2009 WZ104 are accessible via Internet at http://neopage.pochta.ru/ENG/OBSERVS/2009wz104.txt.

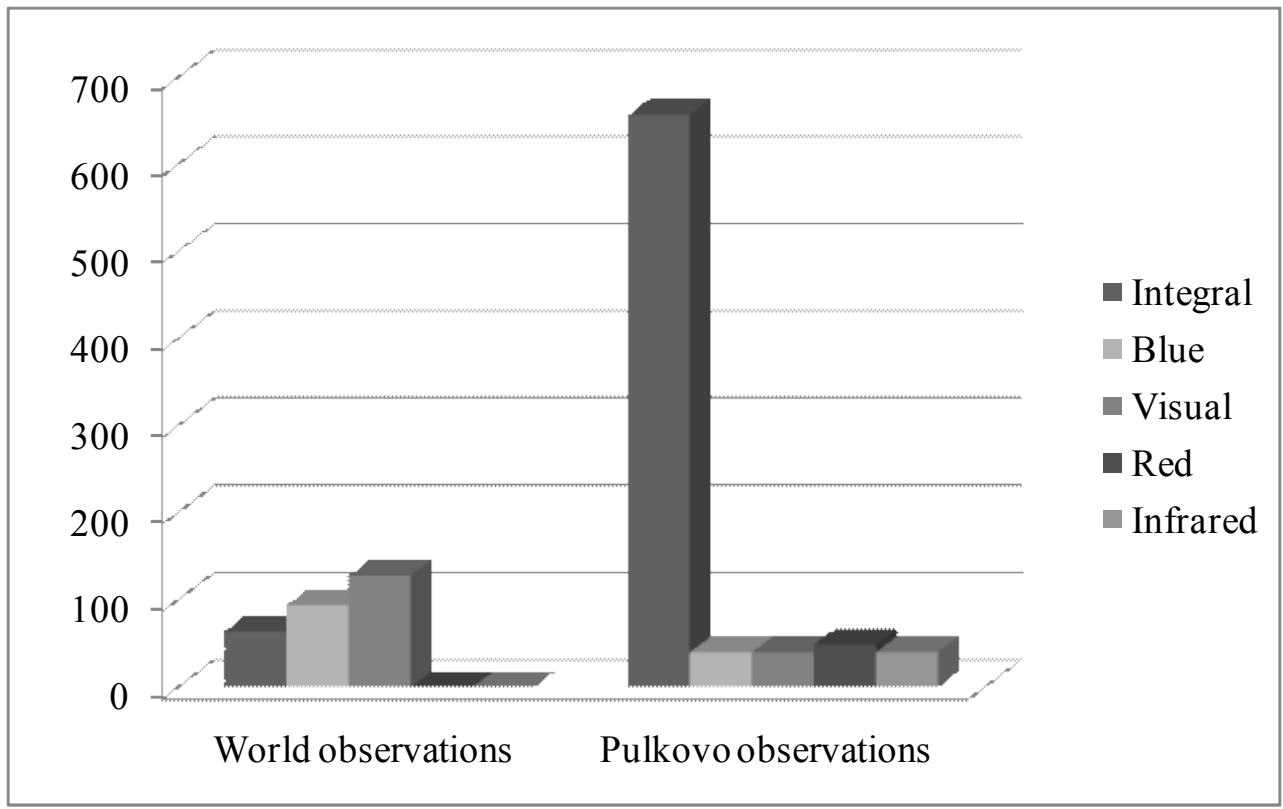

**Fig. 1.** The distribution of the 2009 WZ104 observations.

**Improvement of 2009 WZ104 heliocentric orbit**

For this task, we used the differential method of orbit improvement which is the base of ORBIMPR software developed in Pulkovo observatory. The orbital elements from MPC and Bowell's (ftp://ftp.lowell.edu/pub/elgb/astorb.html) catalogs were used as initial orbit. The observations were processed with use of USNO-B1.0 and UCAC2 catalogs. As one can see in Tables 1 and 2, mean (O–C)s are more precise for UCAC2 catalog both for right ascension and declination. Further analysis is based on the results of processing with UCAC2 catalog.

**Table 1.** Observations with MTM-500M.

|  | $(O-C)_\alpha \cos\delta$, " | $(O-C)_\delta$, " |
|---|---|---|
| USNO-B1.0 | 0.25 ± 0.11 | 0.25 ± 0.11 |
| UCAC2 | 0.18 ± 0.11 | 0.08 ± 0.08 |

**Table 2.** Observations with ZA-320M.

|  | (O-C)$_\alpha$cos$\delta$, " | (O-C)$_\delta$, " |
|---|---|---|
| USNO-B1.0 | 0.47 ± 0.38 | 0.59 ± 0.41 |
| UCAC2 | 0.36 ± 0.32 | 0.35 ± 0.34 |

Then we improve 2009 WZ104 orbit using different sets of observations:
1) 289 observations of different observatories throughout the world (see Table 3);
2) 289 world observations + 174 observations with ZA-320M + 483 observations with MTM-500M (see Table 4).

Improved orbit on the base of world observations and observations of MTM-500M produces smallest deviations, but we shall use orbit improved on the base of world, MTM-500M and ZA-320M observations because this series is more complete.

**Table 3.** The orbit of 2009 WZ104 asteroid improved on the base of the world observations.

| Elements | Initial | Improved |
|---|---|---|
| $M$, ° | 309.064060 | 309.064078 |
| $\omega$, ° | 10.465221 | 10.465231 |
| $\Omega$, ° | 263.350810 | 263.350778 |
| $i$, ° | 9.836916 | 9.836906 |
| $E$ | 0.19245999 | 0.19246003 |
| $a$, a.e. | 0.855704300 | 0.855704363 |
| $q$, a.e. | 0.691015459 | 0.691015473 |

**Table 4.** The orbit of 2009 WZ104 asteroid improved on the base of the world, MTM-500M and ZA-320M observations.

| Elements | Initial | Improved |
|---|---|---|
| $M$, ° | 309.06406 | 309.063995 |
| $\omega$, ° | 10.465221 | 10.465292 |
| $\Omega$, ° | 263.35081 | 263.35076 |
| $i$, ° | 9.836916 | 9.836928 |
| $E$ | 0.19245999 | 0.19246005 |
| $a$, a.e. | 0.8557043 | 0.855704431 |
| $q$, a.e. | 0.691015459 | 0.691015514 |

Coordinates were calculated on the base of elements of both Bowell's catalog and improved orbit. In the last case, the scatter of (O–C) values is smaller and systematic shift disappears. As an example, the (O–C) values for right ascension (reduced to equator) are shown in Fig. 2 (orbit from Bowell's catalog) and Fig. 3 (improved orbit).

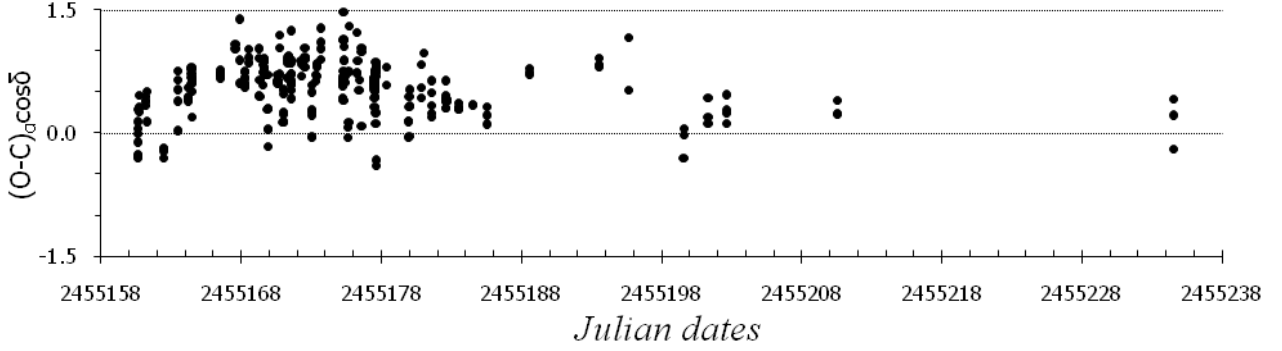

**Fig. 2.** (O-C)$_\alpha$cos$\delta$ for world observations and orbit from Bowell's catalog.

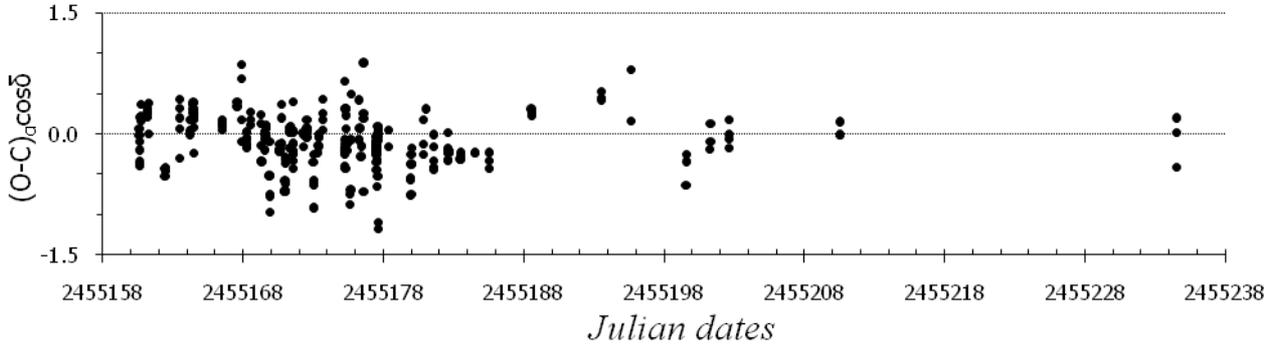

**Fig. 3.** (O-C)$_\alpha$cos$\delta$ for world observations and improved orbit.

**Photometric analysis of the observations**

For the photometric analysis, we have used observations of MTM-500M telescope with *BVRI* filters. Absolute magnitudes of the asteroid in these photometric bands were estimated on the base of the observations. To reduce observed magnitudes to absolute one, we have used equations from (Lagerkvist, Williams, 1987). Due to large error of the slope parameter calculation, we have used its mean value for asteroids of this type: $G = 0.15$ (Vinogradova et al, 2003). The absolute magnitudes for *BVRI* bands are shown in Table 5. Large error for *B* band is due to bad seeing of the asteroid in the filter. The *B–V*, *V–R* and *R–I* color indices were calculated on the base of the absolute magnitudes (see Table 6).

According to absolute magnitudes, the 2009 WZ104 asteroid may belong to R or Q taxonomic class by Tholen (Tholen et al, 1989). The diagram of absolute magnitudes vs. wavelength for 2009 WZ104 and the two classes is shown in Fig. 8. Both of the classes are rare. Their mineralogy is following: chondrites with pyroxene $(Mg,Fe)SiO_3$ and olivine $(Mg,Fe)_2SiO_4$.

**Table 5.** Absolute magnitudes of 2009 WZ104 in *BVRI* bands.

| *B* | *V* | *R* | *I* |
|---|---|---|---|
| $21.17 \pm 0.7^m$ | $20.52 \pm 0.04^m$ | $20.34 \pm 0.04^m$ | $20.60 \pm 0.06^m$ |

**Table 6.** Color indices of 2009 WZ104.

| B–V | V–R | R–I |
|---|---|---|
| 0.65 ± 0.7ᵐ | 0.18 ± 0.05ᵐ | −0.26 ± 0.07ᵐ |

Unfortunately, we cannot precise the classification because there are no any observations at wavelengths longer than 9000 Å for this asteroid throughout the world. Taxonomic class of asteroids is determined by comparing the observed spectral curve with the spectral curve of classes. Searched for the minimum sum of squared deviations between the spectral curves considering the errors.

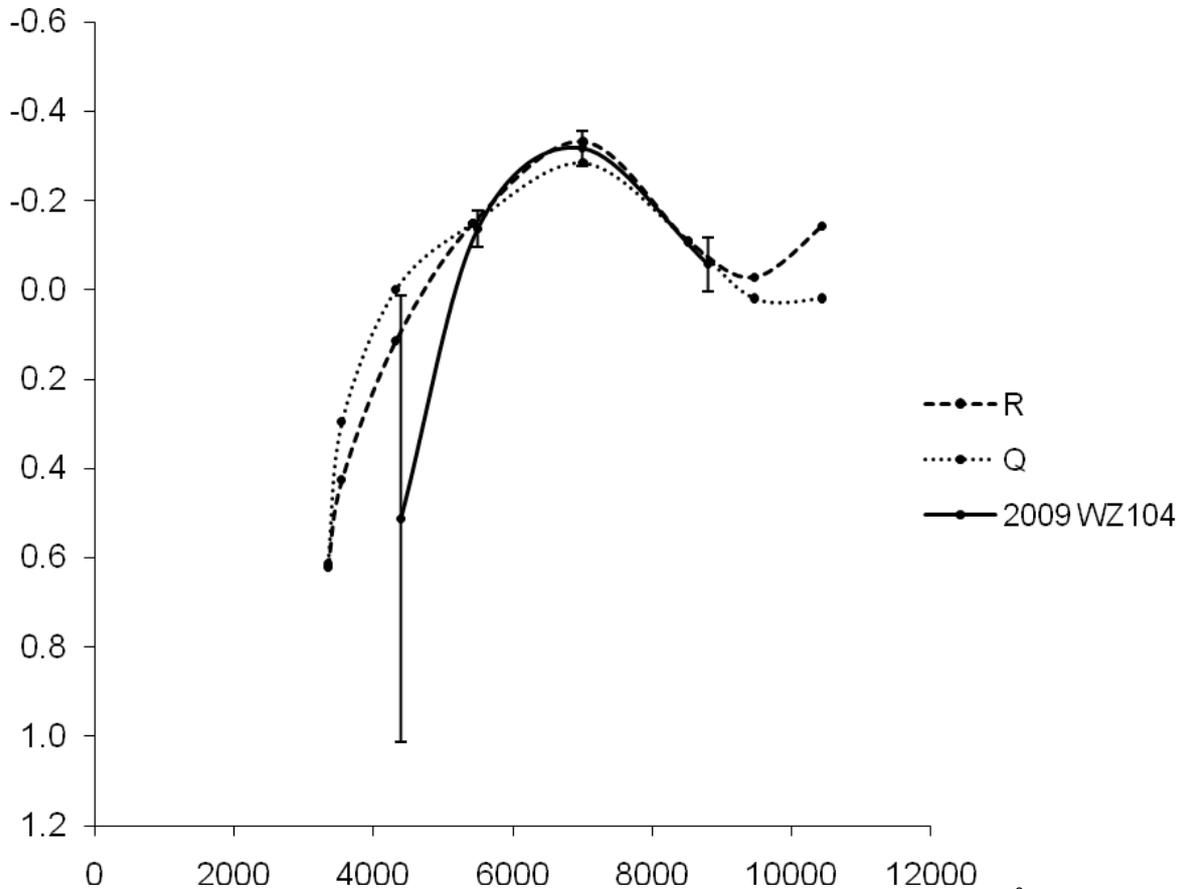

**Fig. 8.** Absolute magnitude of 2009 WZ104 asteroid vs. wavelength (Å) and spectral curves of two taxonomic classes.

**Frequency analysis of 2009 WZ104 lightcurves**

The frequency analysis of 2009 WZ104 observational series was made to search for possible periodicities due to rotation of the asteroid. Two methods were used for the frequency analysis: CLEAN method (Vityazev, 2001a) and Scargle method (Scargle, 1982). The united series of MTM-500M and ZA-320M observations was used for the analysis. Its overall duration is 12 days. During this time, the phase angle of the asteroid changes slightly (2.6°). Hence, we can link the detected periodicity to axial rotation of the asteroid.

The investigated series has low duty cycle. This leads to high noise in Schuster and Scargle periodograms. But these periodograms are very alike: they have peaks at same frequencies and their amplitudes differ very slightly. The periodograms have alike groups of five maxima. It is hard to choose greatest of the maxima because differences of their amplitudes are less than signal-to-noise ratio. Hence, we have five candidates for rotational period. The Table 7 contains frequencies and correspondent possible axial rotation periods for the two methods. But nevertheless, CLEAN method marks one of the peaks and we consider it as most possible although other four should not be discarded.

We have no data of the asteroid form a priori. Hence we can not determine whether the frequency analysis detects full period or half-period. Consequently we can consider two most possible periods of axial rotation of 2009 WZ104: (9.652 ± 0.001) and (19.304 ± 0.002) hours.

**Table 7.** Frequency and periods corresponding to five maxima on Schuster and Scargle periodograms.

| Frequency, day$^{-1}$ | | Period, hours | | |
|---|---|---|---|---|
| *CLEAN* | *Scargle* | *CLEAN* | *Scargle* | *Average* |
| 0.366 | 0.370 | 65.574 | 64.865 | 65.217 ± 0.015 |
| 1.372 | 1.368 | 17.493 | 17.543 | 17.518 ± 0.001 |
| 2.493 | **2.480** | 9.627 | **9.677** | 9.652 ± 0.001 |
| 3.499 | 3.506 | 6.859 | 6.845 | 6.852 ± 0.001 |
| 4.619 | 4.504 | 5.196 | 5.329 | 5.261 ± 0.003 |

**Estimation of dimensions and form of 2009 WZ104 asteroid**

Presuming taxonomic class of 2009 WZ104 as R or Q (see above), we estimate its albedo as 0.3 ÷ 0.4 or 0.16 ÷ 0.21 respectively and mean density as 2.7 g/cm$^3$ (Standish, 2000). Additionally we have considered mean albedo and mean density of NEAs pointed in (Виноградова и др., 2003). As result, we can estimate the diameter of 2009 WZ104 using following formula:

$$D = 1329 \, p_V^{-1/2} \, 10^{-H/5}.$$

Here, $D$ is the diameter, $p_V$ is albedo in $V$ passband, $H$ is absolute magnitude in the same passband. To estimate the mass of the asteroid, we presume that it is spherical. Then we have following formula:

$$m = \pi/6 \, D^3 \, \rho.$$

Here, *m* is mass and ρ is mean density. So we can construct a table (see Table 8) for mass and diameter using mean albedo and density for two taxonomic classes and NEAs (that is usual for poor investigated NEAs). As one can see, both mass and diameter can considerably differ from values in R-class column. In the case of Earth impact, the kinetic energy should be about $10^3$ megatons.

**Table 8.** Physical parameters estimations.

|  | R class | Q class | NEAs |
|---|---|---|---|
| Albedo | 0.3 ÷ 0.4 | 0.16 ÷ 0.21 | 0.154 |
| Mean density | 2.7 g/cm$^3$ | 2.7 g/cm$^3$ | 2.6 g/cm$^3$ |
| Diameter | 165 ÷ 191 m | 227 ÷ 261 m | 267 m |
| Mass | $6.35 \cdot 10^9 \div 9.85 \cdot 10^9$ kg | $1.67 \cdot 10^{10} \div 2.51 \cdot 10^{10}$ kg | $2.59 \cdot 10^{10}$ kg |

**Conclusion**

For the further investigations, it is necessary to get new dense and long series of observations of 2009 WZ104 asteroid during its next approach to Earth in January 2014. It will allow refining its orbit and transferring it to numbered category. For more precise classification, it is necessary to make observations at wavelengths longer than 9000 Å and more precise estimation of absolute magnitude in *B* passband.